\documentclass[12pt]{article}
\usepackage{epsf}

\def\Tr                 {\mathop{\rm Tr}}
\def\none               {\multicolumn{2}{c}{---}}
\def\be{\begin{equation}}
\def\ee{\end{equation}}
\newcommand{\chinfty}{\chi^\infty}
\newcommand{\chiqu}{\chi^{\rm qu}}

\newcommand{\chiquhat}{\hat{\chi}^{\rm qu}}
\def\lsi{\raise0.3ex\hbox{$<$\kern-0.75em\raise-1.1ex\hbox{$\sim$}}}
\def\gsi{\raise0.3ex\hbox{$>$\kern-0.75em\raise-1.1ex\hbox{$\sim$}}}

\newcommand{\gsim}{\mathbin{\gsi}}
\newcommand{\MeV}{\mathop{\rm MeV}}
\newcommand{\fm}{\mathop{\rm fm}}

\begin{document}

\begin{titlepage}


\begin{flushright}
DAMTP-2001-70\\
OUTP-01-43P\\
hep-lat/0108006\\
\hspace{1em}\\
August 2001
\end{flushright} 

\vfill
\begin{centering}

{\bf THE TOPOLOGICAL SUSCEPTIBILITY AND \boldmath{$f_\pi$} \\
FROM LATTICE QCD}
\vspace{0.8cm}

{\sl UKQCD Collaboration}

\vspace{0.8cm}

A. Hart$^{\rm a}$ and 
M. Teper$^{\rm b}$

\vspace{0.3cm}
{\em $^{\rm a}$%
Department of Applied Mathematics and Theoretical Physics, \\
Centre for Mathematical Sciences, University of Cambridge,\\
Wilberforce Road CB3 0WA, UK\\}
\vspace{0.3cm}
{\em $^{\rm b}$%
Theoretical Physics, University of Oxford, 1 Keble Road, \\
Oxford OX1 3NP, UK\\}

\vspace{0.7cm}
{\bf Abstract.}
\end{centering}

\vspace{1ex}
\noindent
We study the topological susceptibility, $\chi$, in QCD with two quark
flavours using lattice field configurations that have been produced 
with an $O(a)$ improved quark action. We find clear evidence for the expected
suppression at small quark mass, and examine the variation of $\chi$
with this mass. The resulting estimate of the pion decay constant,
$f_\pi = 105 \ \pm 6 \ ^{+18}_{-10} \ \MeV$, is consistent with the
experimental value of $\simeq 93 \ \MeV$.  We compare $\chi$ to the
large-$N_c$ prediction and find consistency over a large range of
quark masses. We discuss the benefits of the non--perturbative action 
improvement scheme and of the stategy of keeping the lattice spacing
(nearly) fixed as the quark mass is varied.
We compare our results with other studies and suggest why 
such a quark mass dependence has not always been seen.

\vfill

\end{titlepage}

\section{Introduction}
\label{sec:topol}

In gluodynamics (the pure gauge or ``quenched'' theory) lattice
calculations of the continuum topological susceptibility now appear 
to be relatively free of the systematic errors arising from the 
discretisation, the finite volumes and the various measurement algorithms
employed (for a recent review, see
\cite{Teper:1999wp}).

The inclusion of sea quarks in (``dynamical'') lattice simulations,
even at the relatively large quark masses currently employed, is
numerically extremely expensive, and can only be done for lattices
with relatively few sites (typically $16^3 32$). To avoid significant
finite volume errors, the lattice must then be relatively coarse, 
with, in our case, a spacing $a \simeq 0.1 \ \fm$.  
This is a significant fraction of the mean instanton radius, as
calculated in gluodynamics, and thus precludes a robust, detailed
study of the local topological features of the vacuum in the presence
of sea quarks. The topological susceptibility, on the other hand, may
be calculated with some confidence and provides one of the first
opportunities to test some of the more interesting predictions of
QCD.  Indeed, it is in these measurements that we find some of the
most striking evidence for the presence of the sea quarks (or,
alternatively, for a strong quenching effect) in the lattice
simulations.

We recall that the ensembles used here have been produced with two notable
features
\cite{ukqcd_prog}.
The first is the use of an improved action, such that
leading order lattice discretisation effects are expected to depend
quadratically, rather than linearly, on the lattice spacing (just as
in gluodynamics). In addition, the action parameters have been chosen
to maintain a relatively constant lattice spacing, particularly for
the larger values of the quark mass.

These features have allowed us to see the first clear evidence 
\cite{Hart:2000wr}
for the expected suppression of the topological susceptibility in the
chiral limit, despite our relatively large quark masses. From this
behaviour we can directly estimate the pion decay constant without 
needing to know the lattice operator renormalisation factors that
arise in more conventional calculations.

The structure of this paper is as follows. In Section~\ref{sec_theory}
we discuss the measurement of the topological susceptibility and its
expected behaviour both near the chiral limit, and in the limit of a 
large number of colours, $N_c$. In Section~\ref{sec_meas} we describe 
the UKQCD ensembles and the lattice measurements of the
topological susceptibility over a range of sea quark masses. We fit
these with various ans\"atze motivated by the previous section. We
compare our findings with other recent studies in
Section~\ref{sec_compare}.  Finally, we provide a summary in
Section~\ref{sec_summ}.

These results were presented at the IOP2000 
\cite{Hart:2000wr},
the Confinement IV
\cite{hart_wien}
and, in a much more preliminary form, the Lattice '99
\cite{Hart:1999hy}
conferences. Since then, we have increased the size of several
ensembles and included a new parameter set to try and address the
issue of discretisation effects in our
results. We also have more accurate results from the quenched theory
with which to compare. A brief summary of this work appears in
\cite{ukqcd_prog}.
\section{The topological susceptibility}
\label{sec_theory}

In four--dimensional Euclidean space-time, SU(3) gauge field
configurations can be separated into topological classes, and moving
between different classes is not possible by a smooth deformation of
the fields.  The classes are characterised by an integer--valued
winding number.  This Pontryagin index, or topological charge $Q$, can
be obtained by integrating the local topological charge density
\be
q(x) = \frac{1}{2} \varepsilon_{\mu \nu \sigma \tau}
F^a_{\mu \nu}(x) F^a_{\sigma \tau}(x)
\ee
over all space-time
\be
Q = \frac{1}{32 \pi^2} 
\int q(x) d^4x\, . 
\ee
The topological susceptibility is the
expectation value of the squared charge, normalised by the volume
\be
\chi = \frac{\langle Q^2 \rangle}{V}.
\ee
An isolated topological charge induces an exact zero-mode in
the quark Dirac operator. As a result 
sea quarks in the vacuum induce an instanton--anti-instanton
attraction which becomes stronger as the quark masses, $m_{u}$,
$m_{d}$, \ldots, decrease towards zero (the `chiral limit'),
and the topological charge and susceptibility will be
suppressed to leading order in the quark mass
\cite{DiVecchia:1980ve},
\be
\chi = \Sigma \left( \frac{1}{m_{u}} + \frac{1}{m_{d}} 
 \right)^{-1}
\ee
where 
\be
\Sigma = - \lim_{m_q \to 0} \lim_{V \to \infty} 
\langle 0 | \bar{\psi} \psi | 0 \rangle
\ee
is the chiral condensate (see 
\cite{Leutwyler:1992yt} 
for a recent discussion). Here we have assumed $\langle 0 | \bar{\psi}
\psi | 0 \rangle = \langle 0 | \bar{u} u | 0 \rangle = \langle 0 |
\bar{d} d | 0 \rangle $ and we neglect the contribution of heavier
quarks.  PCAC theory relates this to the pion decay constant $f_\pi$
and mass $M_\pi$ via the Gell-Mann--Oakes--Renner relation as
\be
f_\pi^2 M_\pi^2 = (m_{u} + m_{d}) \Sigma + {\cal O}(m_q^2)
\ee
and we may combine these for $N_f$ degenerate light flavours 
to obtain
\be
\chi = \frac{f_\pi^2 M_\pi^2}{2 N_f} + {\cal O}(M_\pi^4)
\label{eqn_chi_pi2}
\ee
in a convention where the experimental value of $f_\pi \simeq 93 \ 
\MeV$%
\footnote{ 
N.B. there is a common alternative convention, used for internal
consistency in some more preliminary presentations of this data
\cite{ukqcd_prog},
where a factor of 2 is removed from $f_\pi^2$ in
Eqn.~\ref{eqn_chi_pi2}, and where $f_\pi$ is a factor of $\sqrt{2}$
larger, around $132 \ \MeV$.}.
This relation should hold in the limit 
\be
f_\pi^2
M_\pi^2 V \gg 1\, ,
\label{eqn_fpi2_vol}
\ee
which is the Leutwyler--Smilga parameter to leading order in
$M_\pi^2$.  We anticipate our results here to say that even on our
most chiral lattices the {\sc lhs} of Eqn.~\ref{eqn_fpi2_vol} is of
order 10 and so this bound is well satisfied. Thus a calculation of
$\chi$ as a function of $M_\pi$ should allow us to obtain a value of
$f_\pi$.  This method has an advantage over more conventional
calculations in that it does not require us to know lattice operator
renormalisation constants which are required for matching matrix
elements, but which are usually difficult to calculate. In principle,
we require instead knowledge of the renormalisation of the topological
charge operators, but we see in the next Section that this problem can
be readily overcome.

As $m_q$ and $M_\pi$ increase away from zero we expect higher order
terms to check the rate of increase of the topological susceptibility
so that, as $m_q,M_\pi \to \infty$, $\chi$ approaches the quenched value,
$\chiqu$. In fact, as we shall see below, the values of $\chi$ that we
obtain are not very much smaller than $\chiqu$. So there is the danger
of a substantial systematic error in simply applying
Eqn.~\ref{eqn_chi_pi2} at our smallest values of $M_\pi$ in order to
estimate $f_\pi$. To estimate this error it would be useful to have
some understanding of how $\chi$ behaves over the whole range of
$m_q$. This is the question to which we now turn.

There are two quite different reasons why $\chi$ might not be
much smaller than $\chiqu$. The obvious first possibility
is that $m_q$ is large. The second possibility is more
subtle: $m_q$ may be small but QCD may be close to its
large-$N_c$ limit
\cite{tHooft:1974hx}.
Because fermion effects are non-leading in powers of $N_c$, we expect
$\chi \to \chiqu$ for any fixed, non--zero value of $m_q$, however
small, as the number of colours $N_c \to \infty$.  There are
phenomenological reasons
\cite{tHooft:1974hx,Witten:1979kh}
for believing that QCD is `close' to $N_c = \infty$,
and so this is not an unrealistic consideration.
Moreover in the
case of $D=2+1$ gauge theories it has been shown 
\cite{Teper:1998te}
that even SU(2) is close to SU($\infty$). Recent
calculations in four dimensions
\cite{Lucini:2001ej,Teper:1998kw}
indicate that the same is true there. In the present simulations, 
the lighter quark masses straddle the strange quark mass and so it 
is not obvious if we should regard them as being large or small. 
We shall therefore take seriously both the possibilities
discussed above.

We start by assuming the quark mass is small but that we
are close to the large-$N_c$ limit. In this limit, the 
topological susceptibility is known 
\cite{Leutwyler:1992yt} 
to vary as
\be
\chi = \frac{\chinfty M_\pi^2}
{\frac{2 N_f \chinfty}{f_\infty^2} + M_\pi^2}
\label{eqn_nlge_form}
\ee
where $\chinfty$, $f_\infty$ are the quantities at leading order in 
$N_c$. In the 
chiral limit, at fixed $N_c$, this reproduces Eqn.~\ref{eqn_chi_pi2}. 
In the large-$N_c$ limit, at fixed $M_\pi$, it tends to 
the quenched susceptibility $\chinfty$ because 
$f^2_\infty \propto N_c$. The corrections to Eqn.~\ref{eqn_nlge_form}
are of higher order in $M^2_\pi$ and/or lower order in $N_c$.

We now consider the alternative possibility: that $m_q$ is not 
small, that higher-order corrections to $\chi$ will be important 
for most of the values of $m_q$ at which we perform calculations,
and that we therefore need an expression for $\chi$ that
interpolates between $m_q=0$ and $m_q=\infty$. Clearly one cannot 
hope to derive such an expression from first principles, 
so we will simply choose one 
that we can plausibly argue is approximately correct. The
form we choose is 
\be
\chi  =  
\frac{f_{\pi}^2}{ \pi N_f}  M_{\pi}^2
\arctan \left(\frac{ \pi N_f}{f_{\pi}^2} \chiqu
\frac{1}{M_{\pi}^2} \right)
\label{eqn_mlge_form} 
\ee
where $f_{\pi}$ is the pion decay constant in the chiral limit.
The coefficients have been chosen so that this reproduces
Eqn.~\ref{eqn_chi_pi2} when $M_\pi \to 0$ and $\chi \to \chiqu + {\cal
O}(1/M_\pi^4)$ when $M_\pi \to \infty$.  Thus, this interpolation
formula possesses the correct limits and it approaches those limits
with power-like corrections.

We shall use the expressions in
Eqns.~\ref{eqn_chi_pi2},~\ref{eqn_nlge_form} and~\ref{eqn_mlge_form}
to analyse the $m_q$ dependence of our calculated values of $\chi$ and
to obtain a value of $f_\pi$ together with an estimate of the
systematic error on that value. In addition, the comparison with
Eqn.~\ref{eqn_nlge_form} can provide us with some evidence for whether
QCD is close to its large-$N_c$ limit or not.

\section{Lattice measurements}
\label{sec_meas}

We have calculated $\chi$ on five complete ensembles of dynamical
configurations produced by the UKQCD collaboration,
as well as one which is still in progress
\cite{ukqcd_prog}.
Details of these data sets
are given in Table~\ref{tab_ensembles}. The SU(3) gauge fields are
governed by the Wilson plaquette action, with ``clover'' improved
Wilson fermions. The improvement is non--perturbative, with $c_{\rm
sw}$ chosen to render the leading order discretisation errors
quadratic (rather than linear) in the lattice spacing, $a$.

The theory has two coupling constants. In pure gluodynamics the gauge
coupling, $\beta$, controls the lattice spacing, with larger values
reducing $a$ as we move towards the critical value at $\beta = \infty$. 
In simulations with dynamical fermions it has the same role for a fixed
fermion coupling, $\kappa$. The latter controls the quark mass, with
$\kappa \to \kappa_c$ from below corresponding to the massless
limit. In dynamical simulations, however, the fermion coupling also
affects the lattice spacing, which will become larger as $\kappa$ is
reduced (and hence $m_q$ increased) at fixed $\beta$.

The three least chiral UKQCD ensembles (by which we mean largest
$m_\pi / m_\rho$) are e$_6$,~e$_5$ and~e$_4$. By appropriately
decreasing $\beta$ as $\kappa$ is increased, the couplings are
`matched' to maintain a constant lattice spacing
\cite{Irving:1998yu,Garden:1999hs}
%
(which is `equivalent' to $\beta \simeq 5.93$ in gluodynamics with a
Wilson action
\cite{ukqcd_prog})
whilst approaching the chiral limit. The physical
volume and discretisation effects should thus be very similar on these
lattices. The remaining ensembles have lower quark masses, but are at
a slightly reduced lattice spacing. (To have maintained a matched lattice
spacing here would have required reducing $\beta$ to values where
the non--perturbative value $c_{\rm sw}(\beta)$ is not known.)
As the lattices are all $L \gsim
1.5 \fm$, we believe that the minor reduction in the lattice volume
should not lead to significant finite volume corrections.
We also remark that
ensembles~e$_2$ and~e$_3$ have been matched to have approximately
the same chirality, but at (mildly) different lattice spacings.

Four--dimensional lattice theories are scale free, and the
dimensionless lattice quantities must be cast in physical units
through the use of a known scale. For this work, we use the Sommer
scale
\cite{Sommer:1994ce}
%
both to define the lattice spacing for the matching procedure, and to
set the scale. The measured value of $\hat{r}_0$ on each ensemble,
as listed in Table \ref{tab_ensembles},
corresponds to the same physical value of $r_0 = 0.49 \ \fm$.
($\hat{r}_0$ is the dimensionless lattice value of $r_0$ in lattice
units i.e. $\hat{r}_0 = r_0(a)/a$. We use the same notation for other
quantities.)  As we are in the scaling window of the theory, we can
then use the na\"{\i}ve dimensions of the various operators to relate
lattice and physical quantities, e.g.  ${\hat{r}_0}^4 \hat{\chi} = {r_0}^4
\chi + {\cal O}(a^2)$, where we have incorporated the expected
non--perturbative removal of the corrections linear in the lattice
spacing.

Further details of the parameters and the scale determination are 
given in
\cite{ukqcd_prog}.
Measurements were made on ensembles of 400--800 configurations of size
$L^3T = 16^3 32$, separated by ten hybrid Monte Carlo trajectories.
Correlations in the data were managed through jack--knife binning of
the data, using ten bins whose size is large enough that neighbouring
bin averages may be regarded as uncorrelated.

We begin, however, with a discussion of lattice operators and results
in the quenched theory.

\subsection{Lattice operators and \boldmath{$\chiqu$}}
\label{sec_ops}

The simplest lattice topological charge density operator is
\be
\hat{q}(n) = \frac{1}{2} \varepsilon_{\mu \nu \sigma \tau}
\Tr U_{\mu \nu}(n) U_{\sigma \tau}(x)
\label{eqn_qlat_op} 
\ee
where $U_{\mu \nu}(n)$ denotes the product of SU(3) link variables
around a given plaquette. We use a reflection-symmetrised 
version and form
\begin {eqnarray}
\hat{Q} & = & \frac{1}{32\pi^2} \sum_n \hat{q}(n) \, , \\
\hat{\chi} & = & \frac{\langle \hat{Q}^2 \rangle}{L^3 T}\,
\end{eqnarray}
with $a^4(L^3T)$ the lattice volume. In general, $\hat{Q}$ will not
give an integer--valued topological charge due to finite lattice
spacing effects.  There are at least three sources of these. First is
the breaking of scale invariance by the lattice which leads to the
smallest instantons having a suppressed action (at least with the
Wilson action) and a topological charge less than unity (at least with
the operator in Eqn.~\ref{eqn_qlat_op}).  We do not address this
problem in this study, although attempts can be made to correct for it
\cite{Smith:1998wt},
but simply accept this as part of the overall ${\cal O}(a^2)$ error.
In addition to this, the underlying topological 
signal on the lattice is distorted 
by the presence of large amounts of UV noise on the scale of
the lattice spacing
\cite{DiVecchia:1981qi},
and by a multiplicative renormalisation factor
\cite{Campostrini:1988cy}
that is unity in the continuum, but otherwise suppresses the observed
charge. Various solutions to these problems exist
\cite{Teper:1999wp}.
In this study we opt for the `cooling' approach. Cooling explicitly
erases the ultraviolet fluctuations so that the perturbative lattice
renormalisation factors for the topological charge and susceptibility
are driven to their trivial continuum values, leaving ${\cal O}(a^2)$
corrections that may be absorbed into all the other lattice
corrections of this order. We cool by moving through the lattice in a
`staggered' fashion, cooling each link by minimising the Wilson
gauge action applied to each of the three Cabibbo--Marinari SU(2)
subgroups in the link element in turn. (The Wilson gauge action is the
most local, and thus particularly efficient at removing short distance
fluctuations whilst preserving the long range correlations in the
fields.)  Carrying out this procedure once on every link constitutes a
cooling sweep (or `cool'). The violation of the instanton scale
invariance on the lattice, with a Wilson action, is such that an
isolated instanton cooled in this way will slowly shrink, and will
eventually disappear when its core size is of the order of a lattice
spacing, leading to a corresponding jump in the topological charge.
Such events can, of course, be detected by monitoring $\hat{Q}$ as a
function of the number of cooling sweeps, $n_c$.
Instanton--anti-instanton pairs may also annihilate, but this has no
net effect on $\hat{Q}$. However, these observations do motivate us to
perform the minimum number of cools necessary to obtain an estimate of
$\hat{Q}$ that is stable with further increasing $n_c$ (subject to the
above). 

\begin{figure}[tb]
\begin{center}
\leavevmode
\epsfysize=510pt
\epsfbox[20 30 620 730]{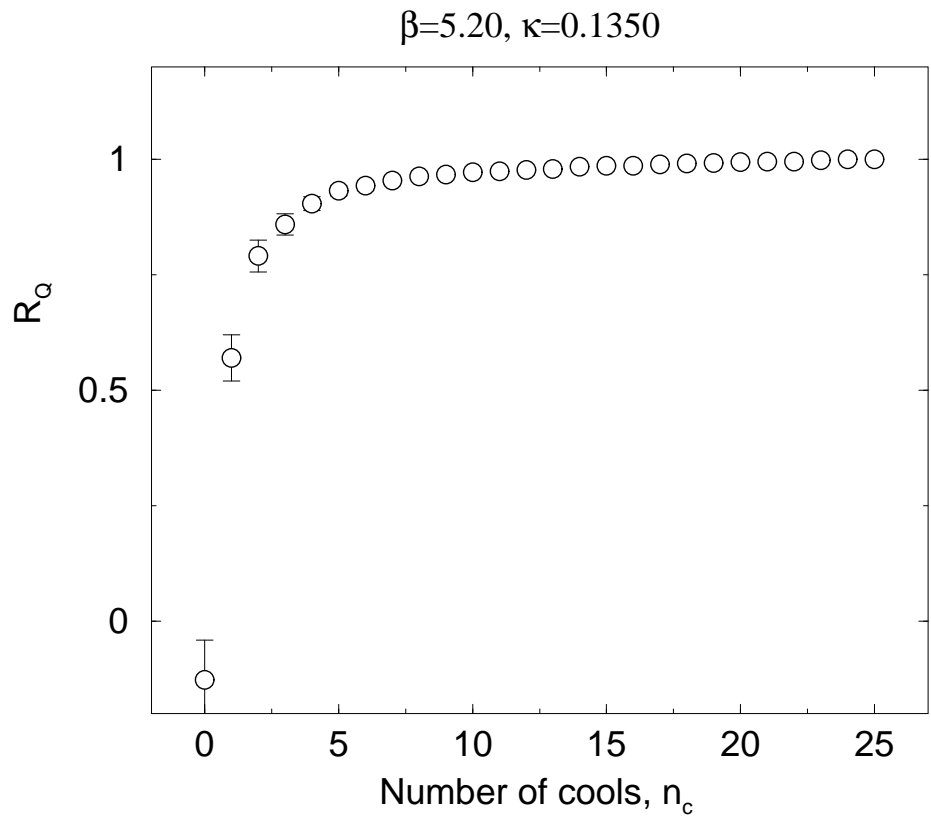}
\end{center}
\caption[]{\label{fig_qcorr}
  { The normalised variation of the topological charge with
    cooling, $R_Q(n_c)$ (Eqn.~\ref{eqn_rq}).}}
\end{figure}

To estimate this point we calculate the normalised correlation
function between the topological charges measured after $n_c$ cooling
sweeps, and a nominally asymptotic 25 cooling sweeps:
\be
R_Q(n_c) = 
\frac{\langle \hat{Q}(n_c)  \hat{Q}(25) \rangle}
{\frac{1}{2}\left( \langle \hat{Q}(n_c)\rangle^2 + 
\langle \hat{Q}(25)\rangle^2 \right)}.
\label{eqn_rq}
\ee
In Fig.~\ref{fig_qcorr} we show a typical plot for ensemble~e$_4$. As
discussed before, we have opted not to attempt to round the
topological charge to integer values. We find $\langle \hat{Q}
\rangle(n_c)$ and $\langle \hat{Q}^2 \rangle(n_c)$ to be stable within
statistical errors for $n_c \gsim 5$, and the results presented here
are for $n_c = 10$.

\begin{figure}[tb]
\begin{center}
\leavevmode
\epsfysize=510pt
\epsfbox[20 30 620 730]{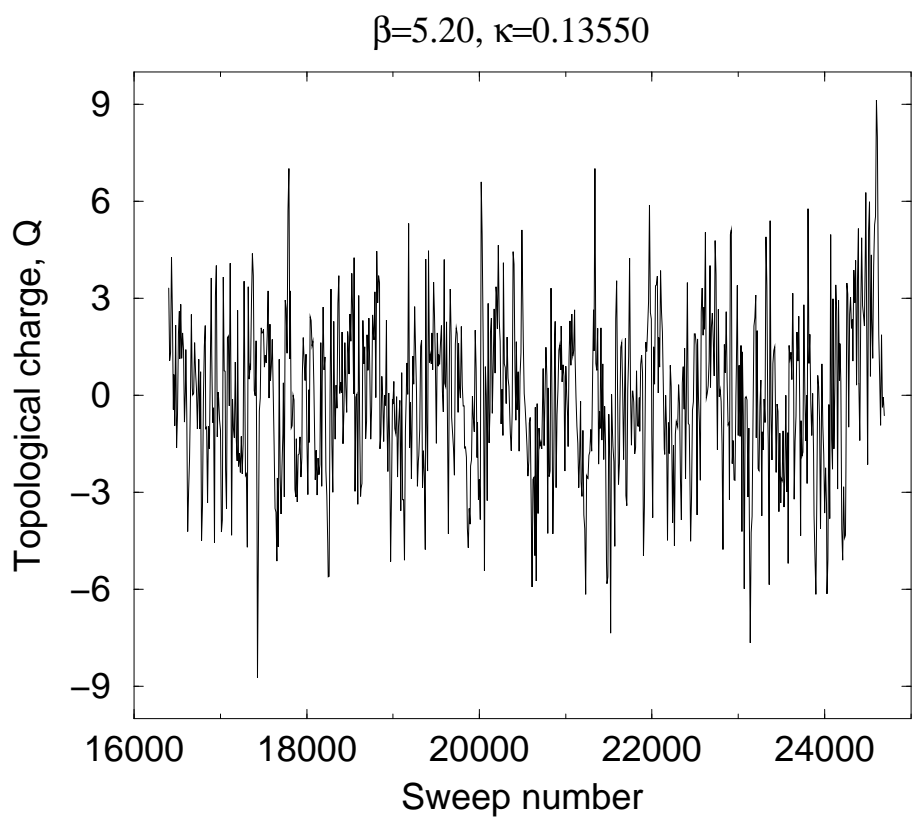}

\end{center}
\caption[]{\label{fig_qseries} { A Monte Carlo time series of
  $\hat{Q}$ after ten cools.}}
\end{figure}

The topological charge of a configuration is related to the smallest
eigenvalues of the Dirac matrix and as such is often believed to be
one of the slowest modes to decorrelate during Monte Carlo
simulations. It is crucial for the error analysis that the bin sizes
for the data are at least twice the integrated autocorrelation
times. In Fig.~\ref{fig_qseries} we plot a typical time series of the
topological charge measured every ten hybrid Monte Carlo
trajectories. The rapid variation between configurations suggests that
the integrated autocorrelation time is small even for the topological
charge.  Estimates of this are given in Table~\ref{tab_chi} in units
of ten trajectories. These reinforce the impression gained from the
time series plots. The bins used in the jack--knife statistical
analysis are between 400 and 1000 trajectories in length and thus may
be confidently assumed to be statistically independent. It is
interesting that although autocorrelation times are hard to estimate
accurately, it does appear that they increase as we move away from the
chiral limit (c.f. Ref.~\cite{ukqcd_prog}).

\begin{figure}[tb]
\begin{center}
\leavevmode
\epsfysize=250pt
\epsfbox[20 30 620 730]{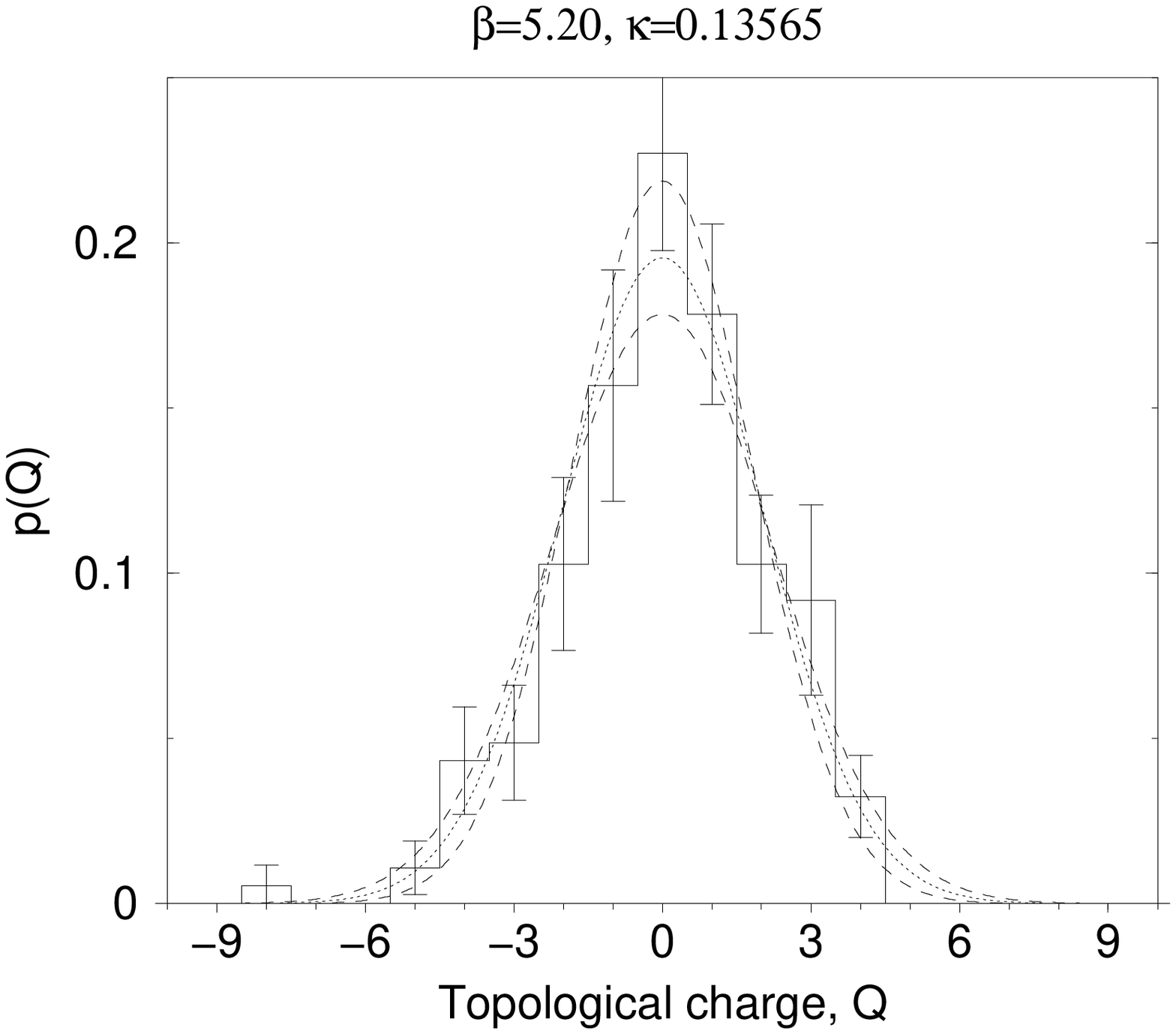}
\epsfysize=250pt
\epsfbox[20 30 620 730]{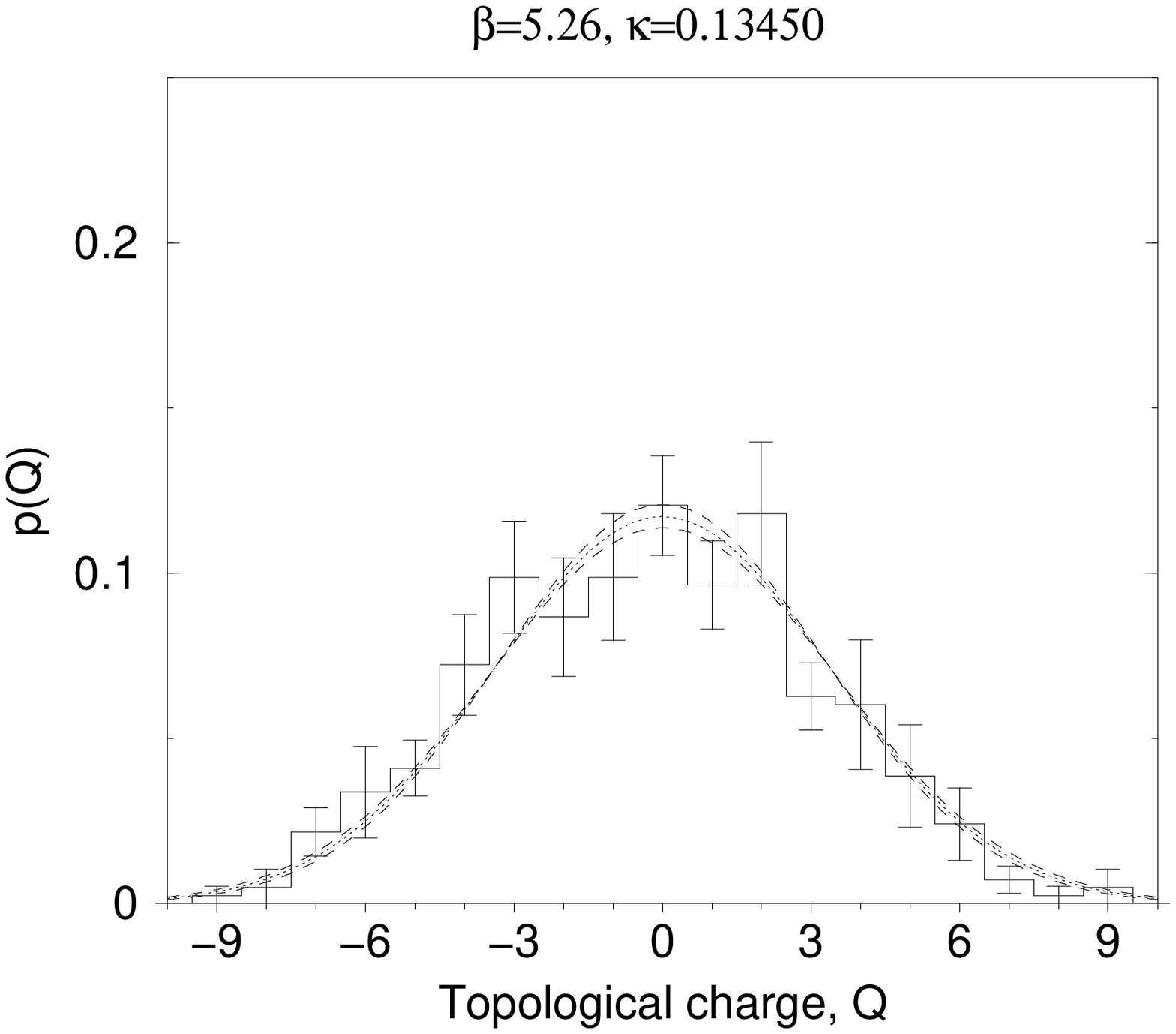}

\end{center}
\caption[]{\label{fig_hist}
  { Histograms of the topological charge for the most and second
    least chiral ensembles,
    together with the Gaussian given in Eqn.~\ref{eq:gaussian}.}}
\end{figure}

In Fig.~\ref{fig_hist} we divide the topological charge measurements
made over an ensemble into bins of unit width centred on the integers,
and plot a histogram of the probability of finding a configuration
with each charge, with errors from the jack--knife analysis. We find
for all our ensembles that these histograms are very close to
being symmetric, centred
around $\hat{Q}=0$ and consistent with a Gaussian envelope. The hybrid
Monte Carlo appears to be sampling the topological sectors correctly,
and it is legitimate to extract an estimate of the topological
susceptibility. On the histograms we show this
estimate as a Gaussian curve
\be
p(\hat{Q}) = \frac{1}{L^3T \hat{\chi}  \sqrt{2\pi}}
\exp -\left[ \frac{\hat{Q}^2}{2L^3 T \hat{\chi}} \right]
\label{eq:gaussian}
\ee
The central line uses our estimate of $\hat{\chi}$, whilst the
outlying curves use the central value plus or minus one standard
deviation. The agreement with the histograms is good.

We also remark that on a lattice one obviously loses instantons with
sizes $\rho \leq {\cal O}(a)$. Since the (pure gauge) instanton
density decreases as $\rho^6$ when $\rho\to 0$ this would appear 
to induce a negligible ${\cal O}(a^7)$ error in the susceptibility. 
However this is only true for $a\to 0$, and the error can be
substantial for the coarse lattices often used in dynamical
simulations.

In general, then, we expect the topological charge and susceptibility
to be suppressed at non--zero lattice spacing. In gluodynamics
with the Wilson action this suppression can, typically, be fitted 
by a leading order, and negative, $O(a^2)$ correction term
starting from quite moderate values of $\beta$. Of course
different ways of calculating the topological charge differ
substantially at finite values of $a$, even if they agree in
the continuum limit. (See, for example, Table 27 in
\cite{Teper:1998kw}.)
An important factor for non--zero $a$ is whether the topological
charge is rounded to the nearest integer after cooling or not, as can
be seen in Table 8 in
\cite{Lucini:2001ej}.
A gluodynamic calculation of $\hat{\chi}$ that uses a method very
similar to the one used in the present paper, in particular an
unrounded topological charge, can be found in the last
column of Table 8 in
\cite{Lucini:2001ej}.
The values of the susceptibility listed there can be
fitted, for $\beta \geq 5.7$, by
\be
{\hat{r}_0}^4 \hat{\chi}^{\rm (qu)} = 0.065 \, (3) - 0.28 \, (4)/{\hat{r}_0}^2
\label{eqn_chiqu_int}.
\ee 
In obtaining this fit we have used the interpolation formula for 
$\hat{r}_0(\beta)$ that is given in 
\cite{Sommer:2001}.
We shall use this formula in Section~\ref{sec_compare}
as a guide to the typical variation of $\hat{\chi}$ with $a$.

In the next Section we shall want to compare our calculations
of $\hat{\chi}$, as obtained in the presence of sea
quarks, with an appropriate quenched limit.
The `equivalent' quenched limit will, of course, depend on
the lattice spacing, i.e. on the value of $\hat{r}_0$. 
However, because of our strategy of varying $\beta$ so as to 
approximately match the values of $\hat{r}_0$ at the 
different values of  $m_q$, the variation in this quenched value,
${\hat{r}_0}^4 \hat{\chi}^{\rm (qu)}$, over the range of lattice
spacings of our ensembles, is in fact much less than the
statistical errors on the measurements themselves.  
(If the calculations had been performed at fixed $\beta$, 
the lattice spacing would have become increasingly coarse
with increasing $m_q$, and the reduction in the quenched
susceptibility would have been much more pronounced over this range of
$\kappa$. We shall return to this important point when we discuss
other work in section~\ref{sec_compare}.) One finds
\cite{ukqcd_prog}
that $\hat{r}_0=4.714 \, (13)$ at $\beta=5.93$ in the quenched theory,
demonstrating that this provides an appropriate quenched limit
for our calculations (see Table \ref{tab_ensembles}). At
$\beta=5.93$ the interpolation formula for $\hat{r}_0(\beta)$
that we used to obtain Eqn.~\ref{eqn_chiqu_int} gives
$\hat{r}_0 \simeq 4.735$ and if we insert this in
Eqn.~\ref{eqn_chiqu_int} we obtain 
${\hat{r}_0}^4 \hat{\chi}^{\rm (qu)} = 0.0525 \, (11)$. Different
methods for calculating $\hat{r}_0$ differ by $O(a^2)$ terms,
and part of the difference between the UKQCD value,
$\hat{r}_0=4.714 \, (13)$, and the Sommer value, $\hat{r}_0 \simeq 4.735$,
might be due to this. If we rescale the susceptibility
to account for this, then the value of
${\hat{r}_0}^4 \hat{\chi}^{\rm (qu)}$ drops to $\simeq 0.0516$.
To do better than this we need to take into account the fact that
the calculations of the topological charge that enter into
Eqn.~\ref{eqn_chiqu_int} are obtained by methods that are
not exactly the same as those used in the calculations of the
present paper. The potentially significant differences are that 
20 cooling sweeps and an unsymmetrised topological charge
were used in
\cite{Lucini:2001ej}.
while we use 10 cooling sweeps and a symmetrised charge.
To estimate the systematic shift induced by these differences
we have performed calculations on 300 $16^4$ lattice
field configurations generated at $\beta=5.93$
(separated by 50 Monte Carlo sweeps). We find
that there is no significant difference between the susceptibility 
as calculated by the symmetrised and unsymmetrised charges,
whether after 10 or 20 cooling sweeps. There is, on the other hand,
a small but significant difference between the susceptibility
as calculated after 10 and 20 cooling sweeps. This reduces
our estimate of the equivalent quenched susceptibility by 
$\sim 0.0025 \, (7)$. So taking all this into account we
take our equivalent quenched susceptibility to be given by
\be
{\hat{r}_0}^4 \hat{\chi}^{\rm (qu)}(\beta=5.93) 
= 0.049 \, (2)
\label{eqn_chiqu_intval}.
\ee 
\subsection{Sea quark effects in the topological susceptibility}

\begin{figure}[tb]
\begin{center}
\leavevmode
\epsfysize=510pt
\epsfbox[20 30 620 730]{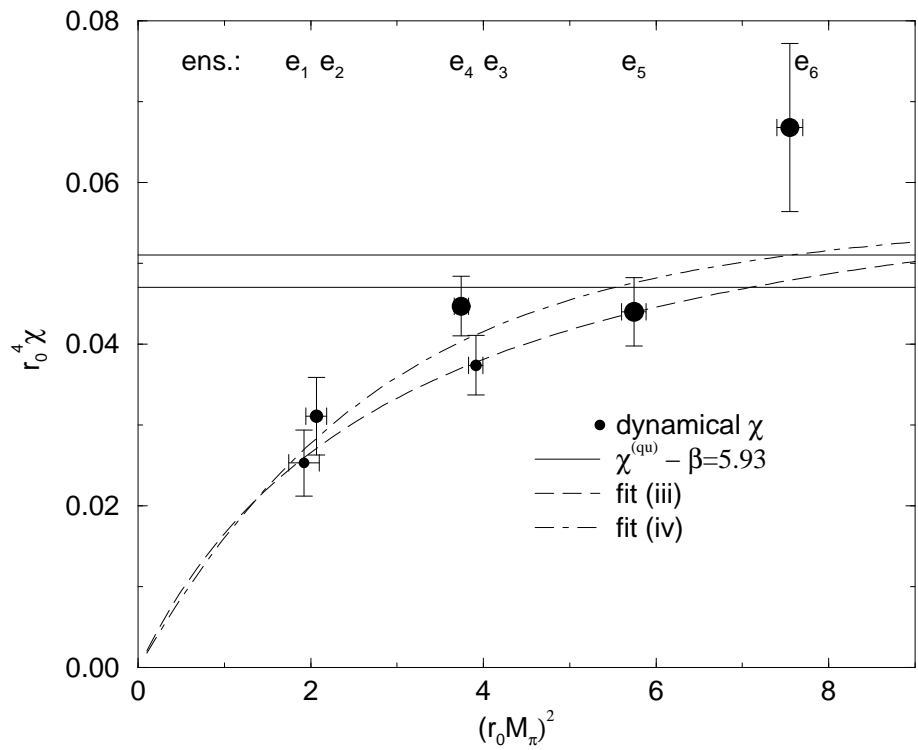}
\end{center}
\caption[]{\label{fig_r04} { The measured topological susceptibility,
    with quenched value at equivalent $\hat{r}_0$. The
    radius of the dynamical plotting points is proportional to
    $\hat{r}_0^{-1}$. The fits, independent of the quenched points,
    are: (iii)~Eqn.~\ref{eqn_fit_nlge} and
    (iv)~Eqn.~\ref{eqn_fit_atan}.}}
\end{figure}

In Table~\ref{tab_chi} we give our estimates of the topological
susceptibility in physical units, using $r_0$ as the scale. In
Fig.~\ref{fig_r04} we plot ${\hat{r}_0}^4\hat{\chi}$ versus a similarly
scaled pseudoscalar meson mass (calculated, of course, with valence
quarks that are degenerate with those in the sea, i.e.  $\kappa_{\rm
valence} = \kappa_{\rm sea}$).  We also plot the corresponding value
of the quenched topological susceptibility, as calculated
at $\beta=5.93$.

Comparing the dynamical and quenched values, the effects of the sea 
quarks are clear. Whilst the measurement on e$_6$ and e$_5$
are consistent with the quenched value, moving to smaller $m_q$
($\propto M_\pi^2$) the topological susceptibility is increasingly
suppressed.  

We can make this observation more quantitative by attempting to fit
our values of ${\hat{r}_0}^4\hat{\chi}$ with the expected functional
form in Eqn.~\ref{eqn_chi_pi2}, so extracting a value of $f_\pi$.  But
we must first be clear whether this fit is justified, and what exactly
we are extrapolating in, bearing in mind that Eqn.~\ref{eqn_chi_pi2}
is strictly a chiral expansion that describes the behaviour for small
sea quark masses in the continuum limit.  

An immediate concern is that 
our cooling technique will occasionally misidentify the value of $Q$
and, in addition, that at finite lattice spacing
the exact zero modes associated with the topological charge
$Q$ are shifted away from zero. All this implies that $\hat{\chi}$ will
not in fact vanish as $m_q \to 0$. However we expect this
effect to be small for the following reasons. First, the use of an 
improved fermion action should ensure that the zero-mode shift will
only be significant for very small instantons, i.e. those whose sizes 
are  $\rho \sim O(a)$.  These are unlikely to survive the cooling
and should not contribute to our calculated value of $Q$. Large
instantons, on the other hand, for which any zero mode shift should 
be negligible, will certainly survive the cooling. The remaining
ambiguity involves the smaller, but not very small, instantons.
These might be erased by the cooling but the probability is small
simply because the number of such charges is small
\cite{Smith:1998wt,Ringwald:1999ze}.
For example, we can see from Fig.~12 in
\cite{Lucini:2001ej}
that the cooling only appears to cut out instantons with 
$\rho \leq 2.5a$. Since the lattice spacing in that plot
is a factor of 1.5 smaller than at our equivalent quenched 
$\beta$ value of 5.93, we would expect that cooling at
$\beta=5.93$ would affect instantons that have $\rho \leq 4a$
in that plot. We see from the SU(3) curve in that figure (after 
scaling up by a factor of $\simeq 4$ to take us from a volume
of $20^4$ at $\beta=6.2$ to a volume of $16^3 32$ at $\beta=5.93$) 
that this involves less than one topological charge per field
configuration. This is a small effect in the present context. 
Thus it is reasonable to assume that this effect can
be neglected for values of $a$ and $m_q$ comparable to the ones that
we study, and that we can then apply Eqn.~\ref{eqn_chi_pi2} to values of
the susceptibility obtained by varying $m_q$ at a fixed value of the
lattice spacing: except that now the decay constant $\hat{f}_\pi$ will
be the one appropriate to that lattice spacing.  

Now, whilst most of
our data points are evaluated on a trajectory of constant lattice
spacing in the parameter space
\cite{Irving:1998yu}, 
not all are. If $\hat{r}_0 \hat{f}_\pi$ varied significantly with $a$
over this range of $a$, it would not be clear how to perform a
consistent chiral extrapolation through the data points.  The
non--perturbative improvement of the action, however, removes the
leading order lattice spacing dependence and the residual corrections
in this range of lattice spacings appear to be small, at least in
measurements of (quenched) hadron spectroscopy
\cite{Wittig:1997bk,Edwards:1997nh}.
An indication of the possible size of the effect on topological
observables comes from comparing our two measurements of ${\hat{r}_0}^4
\hat{\chi}$ at $(\hat{r}_0 \hat{M}_\pi)^2 \simeq 4$. The range of lattice
spacings here ($\hat{r}_0$ varies from 4.75 to 5.14) is comparable to
that over our total data set. The accompanying shift in the
topological susceptiblity is, however, within our statistical
errors. Accordingly, we proceed now to attempt a common chiral
extrapolation to the data, assuming throughout that lattice
corrections to the relations discussed before are too small to be
discernable in our limited data.  We shall at the end of this Section
return to the issue of scaling violations. 

\begin{figure}[tb]
\begin{center}
\leavevmode
\epsfysize=510pt
\epsfbox[20 30 620 730]{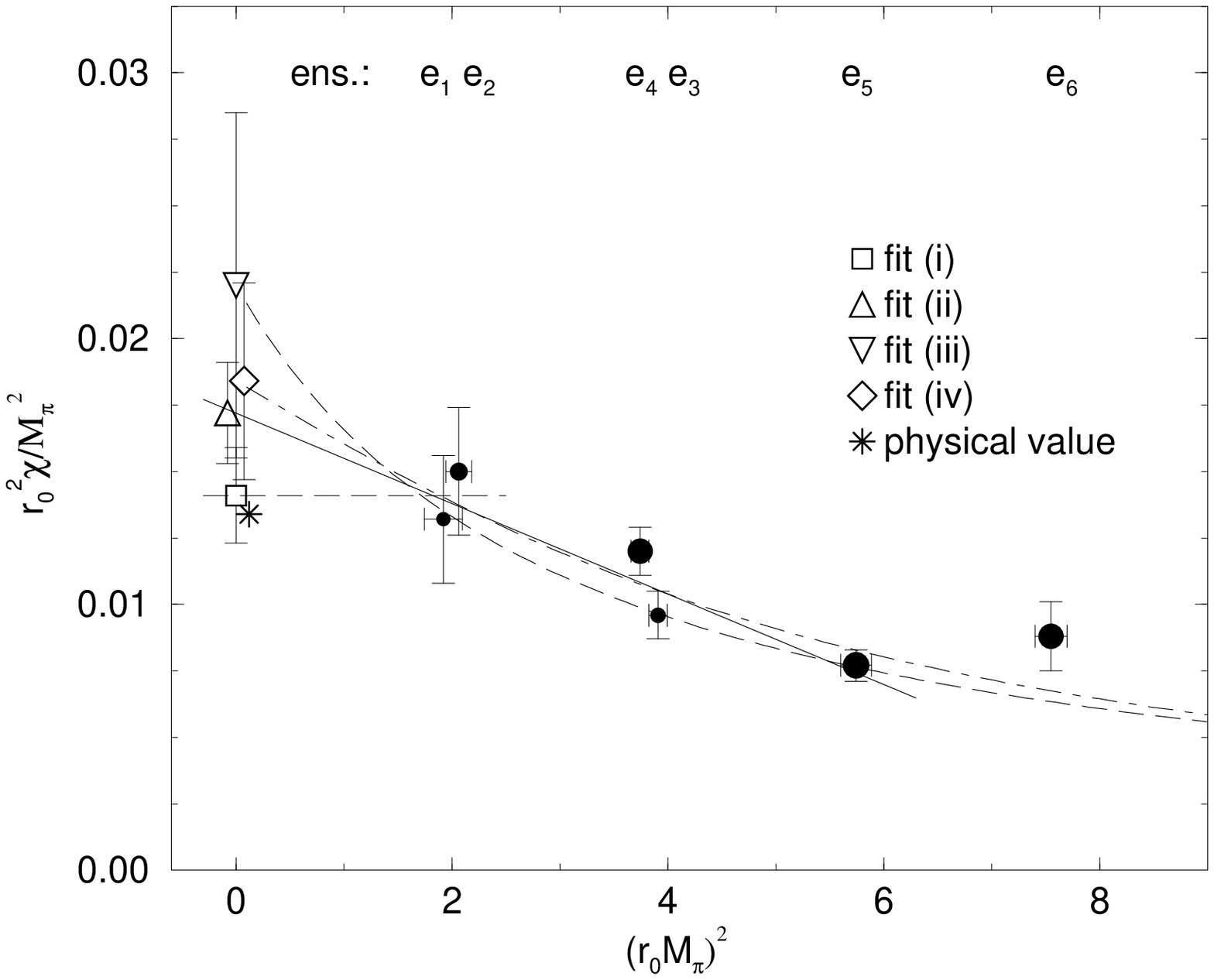}
\end{center}
\caption[]{\label{fig_r02} { The measured topological
    susceptibility. The radius of the dynamical plotting points is
    proportional to $\hat{r}_0^{-1}$. The fits, independent of the
    quenched points, are: (i)~Eqn.~\ref{eqn_fit_fl},
    (ii)~Eqn.~\ref{eqn_fit_fl_lin}, (iii)~Eqn.~\ref{eqn_fit_nlge}
    and (iv)~Eqn.~\ref{eqn_fit_atan}.}}
\end{figure}

For this purpose, and in the light of the discussion at the end of
section~\ref{sec_ops}, it is useful to redisplay the data in
Fig.~\ref{fig_r02}, where the leading order chiral behaviour would
then be a horizontal line,
\be
\frac{{\hat{r}_0}^2\hat{\chi}}{\hat{M}_\pi^2} =
c_0 
\label{eqn_fit_fl} \\
\ee
and including the first correction gives a generic straight line
\be
\frac{{\hat{r}_0}^2\hat{\chi}}{\hat{M}_\pi^2}
 = 
c_0 + c_1 (\hat{r}_0 \hat{M}_\pi)^2.
\label{eqn_fit_fl_lin} 
\ee
In each case the intercept is related to the decay constant by $c_0 =
(\hat{r}_0 \hat{f}_\pi)^2/4$. We now follow a standard fitting
procedure, first using the most chiral points, then systematically
adding the less chiral points until the fit becomes unacceptably
bad. The larger the number of points one can add in this way, the more
evidence one has for the fitted form and the more confident one is
that the systematic errors, associated with the neglected higher order
corrections, are small.  The results of performing such fits are shown
in Table~\ref{tab_fit_res} and those using the two and four most
chiral points respectively are plotted on Fig.~\ref{fig_r02}.  We see
from the Table that the fits using Eqn.~\ref{eqn_fit_fl_lin} show much
greater stability and these are the ones that will provide our
eventual best estimate for $f_\pi$.

We should comment briefly on the determination of the fitting
parameter errors.  In performing all but the constant fit we must contend
with the data having (small) errors on the abscissa in addition to the
ordinate. In order to estimate their affect on the fitting parameters,
we first perform fits to the data assuming that the abscissa data take
their central values. Identical fits are then made using the central
values plus one, and then minus one standard deviation. The spread of
the fit parameters obtained provides what is probably a crude
over--estimate of this error (given there is some correlation between
the ordinal and abscissal uncertainties) but is sufficient to show
that it is minor. We show this spread as a second error, and
for estimates of the decay constant we add it in quadrature to the
other fit parameter error.

It is remarkable that we can obtain stable fits to most of our data
using just the first correction term in Eqn.~\ref{eqn_fit_fl_lin}.
Nonetheless, as we can see in Fig.~\ref{fig_r04}, our values of the
susceptibility are not very much smaller than the $M_\pi = \infty$
quenched value and we need to have some estimate of the possible
systematic errors that may arise from neglecting the higher order
corrections that will eventually check the rise in $\hat{\chi}$. As
discussed earlier we shall do so by exploring two possibilities. One
is that the reason why $\hat{\chi}$ is close to $\chiquhat$ is not
that $m_q$ is `large' but rather that $N_c = 3$ is large.  Then the
values of $\hat{\chi}$ should follow the form in
Eqn.~\ref{eqn_nlge_form}. A second possibility is simply that our
values of $m_q$ are indeed large. In that case we have argued that the
functional form Eqn.~\ref{eqn_mlge_form} should be a reasonable
representation of the true mass dependence. We now perform both types
of fit in turn.

We begin with the first possibility, and therefore fit the data
with the following ansatz
\be
\hat{r_0}^4 \hat{\chi} =  
\frac{c_0 c_3 (\hat{r}_0 \hat{M}_\pi)^2}
{c_3 + c_0 (\hat{r}_0 \hat{M}_\pi)^2},
\label{eqn_fit_nlge}
\ee
where we expect $c_3$ = $\hat{r_0}^4 \chiquhat$ up to ${\cal
  O}(1/N_c^2)$ corrections%
\footnote{For an alternative motivation of this form, see
\cite{Durr:2001ty}.}.
To test this we fit up to seven data points.  The first six are
measured in the dynamical simulations. The final quantity is the
quenched susceptibility at $\beta=5.93$.

We also expect, from the Maclaurin chiral expansion of
Eqn.~\ref{eqn_fit_nlge},
\be \hat{r_0}^4 \hat{\chi} = c_0 \cdot
(\hat{r}_0 \hat{M}_\pi)^2 - {c_0}^2/c_3 \cdot (\hat{r}_0 \hat{M}_\pi)^4
+ c_0^3/c_3^2 \cdot (\hat{r}_0 \hat{M}_\pi)^6 + {\cal O}((\hat{r}_0
\hat{M}_\pi)^8)
\label{eqn_exp_nlge}
\ee
that $c_0$ is related to the decay constant as before, $c_0 =
(\hat{r}_0 \hat{f}_\pi)^2/4$. We present the results of the fits in
Table~\ref{tab_fit_res2}. We find the UKQCD data to be well fitted by
this form, but the asymptotic value is higher than the number we use
for the quenched limit (in contrast to earlier estimates of this
value). Incorporating this number in the fit leads to a poorer
$\chi^2$, and a less robust fit.

We turn now to fits based on the functional form in
Eqn.~\ref{eqn_mlge_form}. We therefore use the ansatz 
\be
\hat{r_0}^4 \hat{\chi}
= 
\frac{2c_0}{\pi} (\hat{r}_0 \hat{M}_\pi)^{2} 
\tan^{-1} \left(
  \frac{c_3}{\frac{2c_0}{\pi}(\hat{r}_0 \hat{M}_\pi)^{2}} \right)
\label{eqn_fit_atan} 
\ee
where once again we expect $c_3$ = $\hat{r_0}^4 \chiquhat$ and
from the expansion
\be
\hat{r_0}^4 \hat{\chi}
 = 
c_0 (\hat{r}_0 \hat{M}_\pi)^2
- \left( \frac{2c_0}{\pi} \right) ^2 \cdot 
\left( \frac{1}{c_3} \right) \cdot (\hat{r}_0 \hat{M}_\pi)^4 
+ {\cal O}((\hat{r}_0 \hat{M}_\pi)^8)
\label{eqn_exp_atan} 
\ee
we expect  $c_0 = (\hat{r}_0 \hat{f}_\pi)^2/4$. Note that in
contrast to Eqn.~\ref{eqn_exp_nlge}, Eqn.~\ref{eqn_exp_atan}
has no term that is cubic in $m_q$ and the rise will remain 
approximately quadratic for a greater range in
$(\hat{r}_0 \hat{M}_\pi)^2$. That this need be no bad thing
is suggested by the relatively large range over which we could fit
Eqn.~\ref{eqn_fit_fl_lin}. Indeed, we see from the fits listed
in Table~\ref{tab_fit_res2} that this form fits our data quite 
well.

Typical examples of the fits from Eqn.~\ref{eqn_fit_nlge} and
Eqn.~\ref{eqn_fit_atan} are shown in Figs.~\ref{fig_r04}
and~\ref{fig_r02}. The similarity of the two functions is apparent. In
Table~\ref{tab_fit_chir} we use the fit parameters to construct the
first three expansion coefficients in the Maclaurin series for the
various fit functions, describing the chiral behaviour of $\chi$. The
fits are consistent with one another.

The fitted asymptote of the susceptibility at large $m_q$ is given by
$c_3$. We see from Table~\ref{tab_fit_res2} that these are broadly
consistent with the quenched value, and our large statistical errors
do not currently allow us to resolve any ${\cal O}(1/N_c^2)$
deviation from this.

As an aside, we ask what happens if we cast aside some of our
theoretical expectations and ask how strong is the evidence from our
data that (a) the dependence is on $M_\pi^2$ rather than on some other
power, and (b) the susceptibility really does go to zero as $M_\pi \to
0$? To answer the first question we perform fits of the kind
Eqn.~\ref{eqn_fit_atan} but replacing $(\hat{r}_0 \hat{M}_\pi)^{2}$ by
$(\hat{r}_0 \hat{M}_\pi)^{c}$. We find, using all seven values of
$\hat{\chi}$, that $c = 1.32 \ (33) \ (10)$; a value  broadly
consistent with $c = 2$. The $\chi^2$/d.o.f. is poorer, however, than
for the fit with a power fixed to 2 (possible as there is one fewer
d.o.f.)  suggesting that the data does not warrant the use of such an
extra parameter. As for the second question, we add a constant
$\hat{c}$ to Eqn.~\ref{eqn_fit_atan} and find $\hat{c} = -0.056 \ (23)
\ (25)$. Again this is consistent with our theoretical expectation;
and again the $\chi^2$/d.o.f. is worse. In both cases, however,
the fits are not robust, with the fit parameters ill-constrained by
our data. (See Table~\ref{tab_fit_res3} for details of the above two
fits.)

Finally, we attempt to address the issue of discretisation
effects. Our use of the non-perturbative $c_{\rm sw}$ should have
eliminated the leading ${\cal O}(a)$ errors. Although $\left( a/r_0
\right)^2$ is small there is, of course, no guarantee that the
coefficient of this correction might not be enhanced. We may begin to
attempt to address this more quantitatively through a combined fit
that includes the first order discretisation correction in
$\hat{r}_0$. Our motivation here is not so much to give a continuum
limit (our data will not really support reliably such a long
extrapolation in $a^2$) as to control the variations due to differing
discretisation in our data over the relatively small range of lattice
spacings in our study. For this reason we fit the simplest combined
ansatz
\be
{\hat{r}_0}^4 \hat{\chi} = c_0 \left( \hat{r}_0 \hat{M}_\pi \right)^2 +
c_1 \left( \hat{r}_0 \hat{M}_\pi \right)^4 + 
\frac{c_2}{{\hat{r}_0}^2}
\ee
to the five most chiral of our data points. The resultant parameters
are shown in Table~\ref{tab_fit_res4}. The change in discretisation
errors across out data set is clearly small, as we expected from the
success of the previous fits.  This justifies the use of a single
chiral extrapolation over this limited range in $\hat{r}_0$. Whilst it
does not, however, rule out deviations between the results of this and
its equivalent in the continuum limit, it does give some indication
that these deviations will be small.  Clearly far greater
accuracy in the measurements is needed to allow a confident
extrapolation to $a = 0$.

Given the consistency of our description of the small $M_\pi$ regime
from our measurements, it is reasonable to use the values of $c_0$ to
estimate the pion decay constant, $f_\pi$. This is done in units of
$r_0$ in Tables~\ref{tab_fit_res} and~\ref{tab_fit_res2}.  We use the
common chiral fit of Eqn.~\ref{eqn_fit_fl_lin} over the largest
acceptable range to provide us with our best estimate and its
statistical error. We then use the fits with other functional forms to
provide us with the systematic error. This produces an estimate
\be
\hat{r}_0 \hat{f}_\pi  =  0.262 \ \pm 0.015 \ ^{+0.046}_{-0.025}
\label{eqn_r0fpi} 
\ee
where the first error is statistical and the second is systematic.
This is of course no more than our best estimate of the value of
$f_\pi$ corresponding to our lattice spacing of $a \simeq 0.1 \fm$.
This value will contain corresponding lattice spacing corrections and
these must be estimated before making a serious comparison with the
experimental value.  We merely note that using $\hat{r}_0 = 0.49 \fm$
we obtain from Eqn.~\ref{eqn_r0fpi} the value
\be
 f_\pi  =  105 \ \pm 6 \ ^{+18}_{-10} \ \MeV
\label{eqn_mevfpi} 
\ee
which is reasonably close to the experimental value
$\simeq 93 \ \MeV$.

\section{Comparison with other studies}
\label{sec_compare}

During the course of this work, there have appeared a number of other 
studies of the topological susceptibility in lattice QCD; in 
particular by the Pisa group
\cite{Alles:1999kf,Alles:2000cg},
the CP-PACS collaboration
\cite{AliKhan:1999zi,AliKhan:2001ym},
the SESAM--T$\chi$L collaboration
\cite{Bali:2001gk}
and the Boulder group
\cite{Hasenfratz:2001wd}.

The most recent studies 
\cite{AliKhan:2001ym,Hasenfratz:2001wd}
are consistent with our findings, but the earlier ones 
found no significant decrease of the susceptibility with
decreasing quark (or pion) mass when everything was expressed in
physical, rather than lattice, units. Indeed when our detailed results 
and analysis were first publicised
\cite{Hart:2000wr}
all the other studies then available 
\cite{Alles:1999kf,Alles:2000cg,AliKhan:1999zi}
(and indeed 
\cite{Bali:2001gk})
appeared to contradict our
findings and it was therefore necessary for us to provide some
reason why this might be so. Although the situation is now different,
the lessons are still useful and we will therefore briefly summarise the
main point here. For more details we refer the reader to
\cite{Hart:2000wr}.

All these other calculations
differ from our study in having been performed at fixed $\beta$. That
implies that the lattice spacing $a$ decreases as $m_q$ is decreased.
In typical current calculations this variation in $a$ is substantial.
(See for example Fig.~4 of
\cite{Allton:1998gi}.)
At the smallest values of $m_q$ the lattice spacing cannot be allowed
to be too fine, because the total spatial volume must remain
adequately large. This implies that at the larger values of $m_q$ the
lattice spacing is quite coarse.  Over such a range of lattice
spacings, the topological susceptibility in the pure gauge theory
typically shows a large variation (as, for example, in
eqn.\ref{eqn_chiqu_int}). Since for coarser $a$ more
instantons (those with $\rho \leq {\cal O}(a)$) are excluded, and more
of those remaining are narrow in lattice units (with a correspondingly
suppressed lattice topological charge) we expect that this variation
is quite general, and not a special feature of the pure gauge theory
with a Wilson action. In lattice QCD, we therefore expect two
simultaneous effects in $\hat{\chi}$ as we decrease $m_q$ at fixed
$\beta$.  First, because of the ${\cal O}(a^2)$ lattice corrections
just discussed, $\hat{\chi}$ will (like $\chiquhat$) tend to increase.
Second, it will tend to decrease because of the physical quark mass
dependence.  In the range of quark masses covered in current
calculations this latter decrease is not very large (as we have seen
in our work) and we suggest that the two effects may largely
compensate each other so as to produce a susceptibility that shows
very little variation with $m_q$, in contrast to the ratio
$\hat{\chi}/\chiquhat$ which does.
 
To illustrate this consider the fixed-$\beta$ calculation in
\cite{Allton:1998gi}.
The range of quark masses covered in that work corresponds to
$(\hat{r}_0 \hat{M}_\pi)^2$ decreasing from about 6.5 to about 3.0.
Simultaneously $a/r_0$ decreases from about 0.437 to about 0.274.
Over this range of $1/\hat{r}_0 \equiv a/r_0$ the pure gauge
susceptibility increases by almost a factor of two, as we see using
Eqn.~\ref{eqn_chiqu_int}. Clearly this is large enough to compensate
for the expected variation of the susceptibility.

As $\beta$ is increased, the $O(a^2)$ variation with $m_q$ of the 
corresponding quenched susceptibility ${\hat{r}_0}^4\chiquhat$
will clearly diminish. Thus we suggested in 
\cite{Hart:2000wr}
that if  CP-PACS were to repeat their susceptibility calculations
on their larger $\beta$ ensembles, they would find an $m_q$
variation of  ${\hat{r}_0}^4\hat{\chi}$ consistent with ours.
This is what they have done in their most recent work
\cite{AliKhan:2001ym}
with the result that we predicted. All this emphasises the utility
of the UKQCD strategy of decoupling the variation of lattice
corrections from the physical $m_q$ dependence, by performing
calculations at fixed $a$ rather than at fixed $\beta$.

\section{Summary}
\label{sec_summ}

We have calculated the topological susceptibility in lattice QCD with
two light quark flavours, using lattice field configurations in which
the lattice spacing is approximately constant as the quark mass is
varied.  We find that there is clear evidence for the expected
suppression of  $\chi$ with decreasing (sea) quark mass.

We have discussed this behaviour in the context of chiral and large $N_c$
expansions, and find good agreement with the functional forms
expected there. We are not able to make a stronger statement
about how close QCD is to its large $N_c$ limit, owing to the 
relatively large statistical errors on our calculated values,
particularly at larger quark masses. This situation should change in the 
near future and, together with the increasing availability of information 
on the large-$N_c$ behaviour of the pure gauge (quenched) theory
\cite{Lucini:2001ej},
a more precise comparison will become possible.

The consistent leading order chiral behaviour from our various fitting
ans\"{a}tze allows us to make an estimate for the pion decay constant,
$f_\pi = 105 \ \pm 6 \ ^{+18}_{-10} \ \MeV$, for the lattice spacing
of $a \simeq 0.1 \fm$. (Here the first error is statistical and the
second has to do with the chiral extrapolation.) We use a lattice
fermion action in which the leading ${\cal O}(a)$ discretisation
errors have been removed. Since the more accurate (quenched) hadron 
masses show little residual lattice spacing dependence
\cite{Wittig:1997bk,Edwards:1997nh},
we might expect that this value of $f_\pi$ 
is close to its continuum limit. In any case, we note
that it is in agreement with the experimental value, $\simeq 93 \
\MeV$.

\section*{Acknowledgments}

The work of A.H. was supported in part by UK PPARC grants
PPA/G/0/1998/00621 and PPA/G/S/1999/00022. A.H. wishes to thank the
Aspen Center for Physics for its hospitality during part of this
work. We thank D. Hepburn and D. Pleiter for preliminary estimates of
lattice pion masses.

%
%
%
%
%

%
%
%
%
%


%
%
%
%
%
\begin{table}[p]
\begin{center}
\begin{tabular}{cccccr@{.}lr@{.}lr@{.}l}
\hline \hline
label &
$\beta$ & $\kappa$ & $c_{\rm sw} $ & $N_{\rm traj.}$ & 
\multicolumn{2}{c}{$\hat{r}_0$} &
\multicolumn{2}{c}{$\hat{m}_\pi/\hat{m}_\rho$} &
\multicolumn{2}{c}{$\hat{r}_0 \hat{m}_\pi$} \\
\hline
e$_1$ & 5.20 & 0.13565 & 2.0171 & 2800 & 5&21 (5)   & \none & 
{\em 1}&{\em 386 (63)} \\
\hline
e$_2$ & 5.20 & 0.13550 & 2.0171 & 8300 & 5&041 (40) & 0&$578^{+13}_{-19}$ &
1&480 (22) \\
e$_3$ & 5.25 & 0.13520 & 1.9603 & 8200 & 5&137 (49) & \none &
{\em 1}&{\em 978 (22)} \\
\hline
e$_4$ & 5.20 & 0.13500 & 2.0171 & 7800 & 4&754 (40) & 0&$700^{+12}_{-10}$ &
1&925 (38) \\
e$_5$ & 5.26 & 0.13450 & 1.9497 & 4100 & 4&708 (52) & 0&$783^{+5}_{-5}$ &
2&406 (27) \\
e$_6$ & 5.29 & 0.13400 & 1.9192 & 3900 & 4&754 (40) & 0&$835^{+7}_{-7}$ &
2&776 (32) \\
\hline \hline
\end{tabular}
\caption{ \label{tab_ensembles} {\em The ensembles studied, with
  measurements made every tenth HMC trajectory. Numbers in italics are
  preliminary estimates.}}
\end{center}
\end{table}
\begin{table}[tb]
\begin{center}
\begin{tabular}{cr@{ }lr@{.}lr@{.}l}
\hline
\hline
ens. & 
\multicolumn{2}{c}{$\tau_{\rm int}(\hat{Q})$} &
\multicolumn{2}{c}{$\langle Q^2 \rangle$} &
\multicolumn{2}{c}{${\hat{r}_0}^4\hat{\chi}$} \\
\hline
e$_1$ & 32& (7) & 4&50 (72) & 
0&0253 (41) \\
e$_2$ & 48& (6) & 6&30 (96) & 0&0311 (48) \\
e$_3$ & \none   & 7&04 (66) & 0&0374 (37) \\
e$_4$ & 79& (5) & 11&48 (90) & 0&0447 (37) \\
e$_5$ & 89& (6) & 11&74 (1.00) & 0&0440 (42) \\
e$_6$ & 129& (19) & 17&13 (2.63) & 0&0668 (104) \\
\hline
$\beta = 5.93$ & \none & 6&47 (24) & 0&049 (2) \\
\hline
\hline
\end{tabular}
\caption{ \label{tab_chi} { The topological charge integrated
  autocorrelation time estimates (in units of HMC trajectories), the
  square of the charge and the susceptibility. The quenched result at
  $\beta=5.93$ is for $16^4$ \cite{Lucini:2001ej}, compared to $16^3
  32$ for the dynamical lattices.}}
\end{center}
\end{table}
\begin{table}[tb]
\begin{center}
\begin{tabular}{ccr@{.}lr@{.}lr@{.}llr@{.}l}
\hline
\hline
Fit & 
$N_{\rm fit}$ &
\multicolumn{2}{c}{$c_0$} &
\multicolumn{2}{c}{$c_1$} &
\multicolumn{2}{c}{$\chi^2/{\rm d.o.f.}$} & $Q$ &
\multicolumn{2}{c}{$\hat{r}_0 \hat{f}_\pi$} \\
\hline
Eqn.~\ref{eqn_fit_fl} & 2 &  
0&0141 (18)     & \none & 0&300 & 0.584 &   0&238 (16) \\
Eqn.~\ref{eqn_fit_fl} & 3 &  
0&0106 (8)      & \none & 2&850 & 0.058 &   0&206 (8)  \\
Eqn.~\ref{eqn_fit_fl} & 4 &  
0&0111 (7)      & \none & 2&311 & 0.074 &   0&211 (7)  \\
Eqn.~\ref{eqn_fit_fl} & 5 &  
0&0096 (5) & \none & 5&289 & 0.000 & \none \\
\hline
Eqn.~\ref{eqn_fit_fl_lin} & 3  & 
0&0187 (36) (5)  & $-0$&0023 (11) (1) & 0&416 & 0.519 & 0&274 (27) \\
Eqn.~\ref{eqn_fit_fl_lin} & 4  & 
0&0184 (37) (5)  & $-0$&0020 (10) (1) & 1&405 & 0.245 & 0&272 (28) \\
Eqn.~\ref{eqn_fit_fl_lin} & 5  & 
0&0172 (19) (1)  & $-0$&0017 (4) (0) & 0&984 & 0.399 & 0&262 (15) \\
Eqn.~\ref{eqn_fit_fl_lin} & 6  & 
0&0149 (16) (1)  & $-0$&0011 (3) (0) & 2&071 & 0.082 & 0&244 (14) \\
\hline
\hline
\end{tabular}
\caption{ \label{tab_fit_res}
  { Fits to the $N_{\rm fit}$ most chiral points of
    $({\hat{r}_0}^2\hat{\chi})/\hat{M}_\pi^2$.}}
\end{center}
\end{table}
\begin{table}[tb]
\begin{center}
\begin{tabular}{ccr@{.}lr@{.}lr@{.}llr@{.}l}
\hline
\hline
Fit & 
$N_{\rm fit}$ &
\multicolumn{2}{c}{$c_0$} &
\multicolumn{2}{c}{$c_3$} &
\multicolumn{2}{c}{$\chi^2/{\rm d.o.f.}$} & $Q$ &
\multicolumn{2}{c}{$\hat{r}_0 \hat{f}_\pi$} \\
\hline
Eqn.~\ref{eqn_fit_nlge} & 3 & 
0&0262 (142) (34) & 0&0592 (252) (35) &  
0&553 & 0.457 & 0&324 (91) \\
Eqn.~\ref{eqn_fit_nlge} & 4 & 
0&0216 (88) (29) & 0&0812 (379) (176) &  
1&441 & 0.237 & 0&294 (64) \\
Eqn.~\ref{eqn_fit_nlge} & 5 & 
0&0257 (86) (24) & 0&0664 (156) (24) &  
1&123 & 0.338 &  0&320 (56) \\
Eqn.~\ref{eqn_fit_nlge} & 6 & 
0&0220 (65) (20) & 0&0773 (202) (37) &  
1&401 & 0.231 & 0&297 (46) \\
Eqn.~\ref{eqn_fit_nlge} & 7 & 
0&0445 (94) (21) & 0&0501 (20) (0) &  
2&134 & 0.261 & 0&422 (46) \\
\hline
Eqn.~\ref{eqn_fit_atan} & 3 & 
0&0210 (69) (21) & 0&0456 (115) (16) &  
0&529 & 0.467 & 0&290 (51) \\
Eqn.~\ref{eqn_fit_atan} & 4 & 
0&0190 (52) (20) & 0&0564 (158) (37) &  
1&408 & 0.245 & 0&276 (41) \\
Eqn.~\ref{eqn_fit_atan} & 5 & 
0&0205 (44) (16) & 0&0517 (73) (12) &  
1&013 & 0.386 &  0&286 (34) \\
Eqn.~\ref{eqn_fit_atan} & 6 & 
0&0184 (37) (15) & 0&0575 (95) (19) & 
1&457 & 0.212 &  0&272 (30) \\
Eqn.~\ref{eqn_fit_atan} & 7 & 
0&0219 (30) (12) & 0&0497 (19) (1) & 
1&408 & 0.300 & 0&296 (22) \\
\hline
\hline
\end{tabular}
\caption{ \label{tab_fit_res2}
  { Fits to the $N_{\rm fit}$ most chiral points of
    $({\hat{r}_0}^4\hat{\chi})$.}}
\end{center}
\end{table}
\begin{table}[tb]
\begin{center}
\begin{tabular}{ccr@{.}lr@{.}lr@{.}l}
\hline
\hline
Fit & 
$N_{\rm fit}$ &
\multicolumn{2}{c}{const.} &
\multicolumn{2}{c}{${\cal O}((\hat{r}_0\hat{M}_\pi)^2)$} &
\multicolumn{2}{c}{${\cal O}((\hat{r}_0\hat{M}_\pi)^4)$} \\
\hline
Eqn.~\ref{eqn_fit_fl_lin} & 4 &
0&0172 (20) & $-0$&0017 (4) & \multicolumn{2}{c}{---}  \\
\hline
Eqn.~\ref{eqn_fit_nlge} & 5 &
0&0220 (68) & $-0$&0063 (43) & 0&0018 (20) \\
Eqn.~\ref{eqn_fit_nlge} & 6 &
0&0286 (39) & $-0$&0130 (35) & 0&0059 (24) \\
\hline
Eqn.~\ref{eqn_fit_atan} & 5 &
0&0184 (40) & $-0$&0024 (12) & \multicolumn{2}{l}{0}  \\
Eqn.~\ref{eqn_fit_atan} & 6 &
0&0170 (18) & $-0$&0019 (4) & \multicolumn{2}{l}{0}  \\
\hline
\hline
\end{tabular}
\caption{ \label{tab_fit_chir}
{ Chiral expansion terms of fitted functions.}}
\end{center}
\end{table}
\begin{table}[tb]
\begin{center}
\begin{tabular}{cr@{.}lr@{.}lr@{.}lr@{.}lr@{.}ll}
\hline
\hline
$N_{\rm fit}$ &
\multicolumn{2}{c}{$c_0$} &
\multicolumn{2}{c}{$c_3$} &
\multicolumn{2}{c}{$c$} &
\multicolumn{2}{c}{$\hat{c}$} &
\multicolumn{2}{c}{$\chi^2/{\rm d.o.f.}$} & $Q$ \\
\hline
7 & 
0&086 (32) (45) & 0&105 (23) (35) &  \multicolumn{2}{c}{\rm fixed to 2}  & 
$-0$&056 (23) (35) &
1&570 & 0.180 \\
7 & 
0&0160 (53) (26) & 0&0492 (20) (0) & 1&32 (33) (10)  &  
\multicolumn{2}{c}{\rm fixed to 0}  &
1&573 & 0.178 \\
\hline
\hline
\end{tabular}
\caption{ \label{tab_fit_res3}
{ Fits to the $N_{\rm fit}$ most chiral values  of
$({\hat{r}_0}^4\hat{\chi})$;
for the power $c$ of $M^2_\pi$, and for the chiral intercept,
$\hat{c}$. These are as described in the text.}}
\end{center}
\end{table}
\begin{table}[tb]
\begin{center}
\begin{tabular}{cr@{.}lr@{.}lr@{.}lr@{.}ll}
\hline
\hline
$N_{\rm fit}$ &
\multicolumn{2}{c}{$c_0$} &
\multicolumn{2}{c}{$c_1$} &
\multicolumn{2}{c}{$c_2$} &
\multicolumn{2}{c}{$\chi^2/{\rm d.o.f.}$} & $Q$ \\
\hline
5 & 
0&0148 (75) & $-0$&0014 (10) &  
0&0055 (120) &
1&611 & 0.200 \\
\hline
\hline
\end{tabular}
\caption{ \label{tab_fit_res4} { Simultaneous chiral and
lattice spacing correction fit to the $N_{\rm fit}$ most chiral values
of $({\hat{r}_0}^4\hat{\chi})$, as described in the text.}}
\end{center}
\end{table}

\end{document}